\documentclass[aps,nofootinbib,superscriptaddress,11pt]{revtex4-1}
\usepackage[letterpaper,hmargin=1in,vmargin=1in]{geometry}

\usepackage{graphicx,epstopdf,amsmath,amsfonts,amssymb,amsxtra,color,phonetic}

\usepackage{multirow,array,booktabs}
\usepackage{cancel,bbold,extarrows,mathdots}
\usepackage{hyperref,cleveref}

\begin{document}
 
\title{The extended Kerr-Schild approach to general relativity}

\author{Xun Wang}
\affiliation{\it MoE Key Laboratory of Fundamental Quantities Measurement and School of Physics,\\
	Huazhong University of Science and Technology,\\
	Luoyu Lu 1037, Wuhan 430074, China}
\email{xungravity@gmail.com}
\author{Jianwei Mei}
\affiliation{\it TianQin Research Center for Gravitational Physics and School of Physics and Astronomy,\\
Sun Yat-Sen University, SYSU Zhuhai Campus, Zhuhai 519082, China}
\email{meijw@sysu.edu.cn}

\date{\today}

\begin{abstract}
	
\noindent
We study in some detail the ``extended Kerr-Schild'' formulation of general relativity, which decomposes the gauge-independent degrees of freedom of a generic metric into two arbitrary functions and the choice of a flat background tetrad. We recast Einstein's equations and spacetime curvatures in the extended Kerr-Schild form and discuss their properties, illustrated with simple examples.
\end{abstract}
\maketitle


\section{Introduction}

The Kerr-Schild (KS) type metric \cite{Trautman:1962,KS,Xanthopoulos} represent an important and interesting class of solutions of Einstein's equations. It establishes the equality of a one-parameter family of exact solutions $\tilde g(\lambda)$ to the linearized solution $d\tilde g(\lambda)/d\lambda\rvert_{\lambda=0}$, with the metric written as $\tilde g_{\mu\nu}=g_{\mu\nu}+\lambda k_\mu k_\nu$ for some null vector field $k$. The full Ricci tensor in fact contains only up to $\lambda^2$ terms and can be shown to vanishes identically if the linearized equation is formally satisfied ($\lambda$ need not to be small). Investigations on various generalizations of the KS ansatz have been fruitful. Recently, it was shown \cite{NotSo} that the nice feature that Einstein's equations truncate at finite order is inherited by the ``extended Kerr-Schild'' (xKS) ansatz $\tilde g_{\mu\nu}=W^2(g_{\mu\nu}+hl_{(\mu} n_{\nu)})$ for a pair of null vectors $l,n$. Any generic metric can be written in the xKS form as a result of the \emph{flat deformation theorem} \cite{Llosa&Soler,Llosa:2008}. It is desirable and the subject of this work to translate basic ingredients of general relativity into the xKS form, serving as a basis for further applications.

We start by detailing the setup of xKS ansatz, using the Newman-Penrose (NP) null tetrad formalism. Then we calculate the Riemann tensor and from it the Ricci and Weyl tensors. In presenting the results, we emphasize on the clear separation between the two functions $W,h$ and the background tetrad (their spin coefficients) as dictated by the xKS ansatz. The classification of algebraic types of the full spacetime is illustrated with special examples. We conclude this paper after briefly comparing xKS form with the double-null foliation.

\section{\lowercase{x}KS form in NP formalism}

\subsection{The metric ansatz}

It is convenient to present the xKS formulation using the NP formalism, in which a tetrad (set of four null vectors) $\{e_1=l,e_2=n,e_3=m,e_4=\bar m\}$ is introduced such that
\begin{equation}
e_a^\mu{e_b}_\mu=\eta_{ab}=
\begin{pmatrix}
0&-1&0&0\\-1&0&0&0\\0&0&0&1\\0&0&1&0
\end{pmatrix}
=\eta^{ab},\quad g_{\mu\nu}=\eta^{ab}{e_a}_\mu{e_b}_\nu
\end{equation}
where lower case Latin letters $a,b=1,2,3,4$ are tetrad indices and bar denotes complex conjugate. Compared to the traditional NP formalism, we use the signature $(-+++)$ (as in \cite{ExactSolns}) in stead of $(+---)$. We assume the same contravariant components of tetrad vectors for both signature conventions (e.g., $l^t,l^r>0$ for $l$ being future-pointing and outgoing), and add an extra sign to covariant components. 

The `extended Kerr-Schild' (xKS) ansatz is the writing of a generic metric $\tilde g$, according to the \emph{flat deformation theorem} \cite{Llosa&Soler,Llosa:2008}, as
\begin{align}
\tilde g_{\mu\nu}&=W^2\bigl[-(1+h/2)(l_\mu n_\nu+l_\nu n_\mu)+m_\mu\bar m_\nu+m_\nu\bar m_\mu\bigr]\label{tilde_gdn}\\
\tilde g^{\mu\nu}&=W^{-2}\bigl[-(1+h/2)^{-1}(l^\mu n^\nu+l^\nu n^\mu)+m^\mu\bar m^\nu+m^\nu\bar m^\mu\bigr]\label{tilde_gup}
\end{align}
where $[l,n,m,\bar m]$ is the tetrad for a \emph{flat} background metric $g$ and $W,h$ are functions. The tetrad $\tilde e_a$ for the full metric $\tilde g$ can be given as
\begin{equation}\label{fulltetrad}
[\tilde l_\mu,\tilde n_\mu,\tilde m_\mu]=[\chi l_\mu,\chi n_\mu,Wm_\mu],\quad[\tilde l^\mu,\tilde n^\mu,\tilde m^\mu]=[\chi^{-1}l^\mu,\chi^{-1}n^\mu,W^{-1}m^\mu],\quad\chi\equiv W(1+h/2)^{\frac{1}{2}}.
\end{equation}
From now on tensorial quantities without tilde are reserved for the background spacetime. To raise and lower indices, one uses $\tilde g$ or $g$ according to whether the quantity has tilde or not.

A quick example is given by the Schwarzschild metric (see also \cite{Llosa:2008}):
\begin{equation}\begin{aligned}
ds^2&=-fdt^2+f^{-1}dr^2+r^2d\Omega^2\\
&=f(-dt^2+dr_*^2)+\frac{r^2}{r_*^2}r_*^2d\Omega^2,\quad(dr_*=f^{-1}dr,f=1-2m/r).
\end{aligned}\end{equation}
Then by writing $-dt^2+dr_*^2=-2l_\mu n_\nu dx^\mu dx^\nu,r_*^2d\Omega^2=2m_\mu\bar m_\nu dx^\mu dx^\nu$, one identifies $\chi^2=f,W^2=r^2/r_*^2$. Note that Schwarzschild metric in Kruskal coordinates is also of xKS form.

\subsection{Riemann, Ricci and Weyl tensors}

It is interesting to see if the xKS ansatz will provide useful ways to find exact solutions of Einstein equations and to explore spacetime properties. As a first step to this end, we recast Riemann, Ricci and Weyl tensors in terms of $W,h$ and the background tetrad. Riemann tensor can be written in the `background+difference' form as \cite{Wald}
\begin{equation}\label{fullRiem}
\tilde R^\rho_{\sigma\mu\nu}=R^\rho_{\sigma\mu\nu}+\mathcal R^\rho_{\sigma\mu\nu}=R^\rho_{\sigma\mu\nu}+\nabla_\mu K^\rho_{\sigma\nu}-\nabla_\nu K^\rho_{\sigma\mu}+K^\rho_{\mu\lambda}K^\lambda_{\sigma\nu}-K^\rho_{\nu\lambda}K^\lambda_{\sigma\mu},
\end{equation}
with
\begin{equation}\label{K}
K^\rho_{\sigma\nu}\equiv\tilde\Gamma^\rho_{\sigma\nu}-\Gamma^\rho_{\sigma\nu}=\frac{1}{2}\tilde g^{\rho\alpha}(\nabla_\nu\tilde g_{\alpha\sigma}+\nabla_\sigma\tilde g_{\alpha\nu}-\nabla_\alpha\tilde g_{\sigma\nu}),
\end{equation}
where $\tilde g,\tilde R,\tilde\Gamma$ are for the full spacetime, $R,\Gamma,\nabla$ are for the background spacetime, and $\mathcal{R}$ is the `difference'. We set $R^\rho_{\sigma\mu\nu}=0$ for flat background. The tetrad components of $\tilde R^\rho_{\sigma\mu\nu}$ w.r.t the full tetrad is thus 
\begin{equation}
\tilde R_{abcd}=W^p\chi^q\mathcal R_{abcd},
\end{equation}
where on the r.h.s we factor out powers of $W,\chi$ by transforming to the flat tetrad via \eqref{fulltetrad}, namely, we define $\tilde R_{abcd}\equiv\tilde R^\rho_{\sigma\mu\nu}{{}\tilde e_a}_\rho\tilde e_b^\sigma\tilde e_c^\mu\tilde e_d^\nu,\mathcal R_{abcd}\equiv\mathcal R^\rho_{\sigma\mu\nu}{e_a}_\rho e_b^\sigma e_c^\mu e_d^\nu$. E.g., for $\tilde R_{1234}$, $W^p\chi^q=\chi\chi^{-1}W^{-1}W^{-1}=W^{-2}$. Note that $\mathcal R_{abcd}$ does not possess the symmetries of Riemann tensor, e.g., $\mathcal R_{3141}=(1+h/2)\mathcal R_{1314}$.

In terms of $\mathcal R_{abcd}$, we have for $\tilde R_{ac}\equiv\eta^{bd}\tilde R_{abcd}$
\begin{equation}\begin{aligned}\label{Ricci_Riem}
\tilde R_{11}&=2W^{-2}\mathcal R_{1314},\quad\tilde R_{22}=\tilde R_{11}(l\leftrightarrow n),\quad\tilde R_{33}=-2W^{-2}\mathcal R_{1323}\\
\tilde R_{13}&=W^{-1}\chi^{-1}\mathcal R_{1213}+W^{-3}\chi\mathcal R_{1334},\quad\tilde R_{23}=\tilde R_{13}(l\leftrightarrow n)\\
\tilde R_{12}&=W^{-2}(2\mathcal R_{1324}-\mathcal R_{1234})+\chi^{-2}\mathcal R_{1212}\\
\tilde R_{34}&=-W^{-2}(2\mathcal R_{1324}-\mathcal R_{1234}+\mathcal R_{3434})\\
\tilde R&=2(\tilde R_{34}-\tilde R_{12}),
\end{aligned}\end{equation}
up to complex conjugation which switches indices $3\leftrightarrow4$, and also using the cyclic symmetry of Riemann tensor. The Weyl scalars are given as
\begin{equation}\begin{aligned}\label{Psi_R}
\tilde\Psi_0&=W^{-2}\mathcal R_{1313},\quad\bar{\tilde\Psi}_4=\tilde\Psi_0(l\leftrightarrow n)\\
\tilde\Psi_1&=\frac{1}{2}\bigl(W^{-1}\chi^{-1}\mathcal R_{1213}-W^{-3}\chi\mathcal R_{1334}\bigr),\quad\bar{\tilde\Psi}_3=\tilde\Psi_1(l\leftrightarrow n)\\
\tilde\Psi_2&=-\frac{1}{3}W^{-2}\Bigl[\mathcal R_{1234}+\mathcal R_{1324}-\frac{1}{2}\mathcal R_{3434}-\frac{\mathcal R_{1212}}{2+h}\Bigr].
\end{aligned}\end{equation}
The basic quantities $\mathcal R_{abcd}$ can be calculated by substituting the xKS form of $\tilde g$ into \eqref{K} and \eqref{fullRiem}. We systematically apply the following rules along the way:
\begin{gather}
e_a^\mu{e_b}_\mu=\eta_{ab},\quad l^\mu\nabla_\mu=\nabla_l,\;l^\mu n^\nu\nabla_\mu\nabla_\nu=\nabla_{ln},\text{ etc.},\quad e_c^\mu{e_a}_{\mu;\nu}e_b^\nu=-\gamma_{cab},
\end{gather}
where $\gamma_{cab}$ (minus sign for our signature convention) are spin coefficients with designated symbols:
\begin{align}
\kappa&=\gamma_{311} & \rho&=\gamma_{314} & 
\sigma&=\gamma_{313} & \mu&=\gamma_{243} & \varepsilon&=\frac{1}{2}\left(\gamma_{211}+\gamma_{341}\right) & \gamma&=\frac{1}{2}\left(\gamma_{212}+\gamma_{342}\right)\\
\lambda&=\gamma_{244} & \tau&=\gamma_{312} &
\nu&=\gamma_{242} & \pi&=\gamma_{241} & \alpha&=\frac{1}{2}\left(\gamma_{214}+\gamma_{344}\right)& \beta&=\frac{1}{2}\left(\gamma_{213}+\gamma_{343}\right).
\end{align}
Under $l\leftrightarrow n$, i.e., switching of indices $1\leftrightarrow2$, one has
\begin{equation}\label{l<->n}
\kappa\leftrightarrow-\bar\nu,\;\rho\leftrightarrow-\bar\mu,\;\sigma\leftrightarrow-\bar\lambda,\;\alpha\leftrightarrow-\bar\beta,\;\epsilon\leftrightarrow-\bar\gamma,\;\pi\leftrightarrow-\bar\tau.
\end{equation}

Now, defining $W^2\equiv1+\omega$, we have explicitly,
\begin{multline}
\mathcal R_{1212}=\\
-\frac{\nabla_{ln}\omega}{1+\omega}-\frac{\nabla_{ln}h}{2+h}+\frac{2+h}{4(1+\omega)^2}\nabla_m\omega\nabla_{\bar m}\omega+\frac{\nabla_m\omega\nabla_{\bar m}h+c.c.}{4(1+\omega)}+\frac{\nabla_mh\nabla_{\bar m}h}{4(2+h)}+\frac{\nabla_l\omega\nabla_n\omega}{(1+\omega)^2}+\frac{\nabla_lh\nabla_nh}{(2+h)^2}\\
+\frac{h}{4}\bigl[(\pi-\bar\tau)\bigl(\frac{\nabla_m\omega}{1+\omega}+\frac{\nabla_mh}{2+h}\bigr)+2\kappa\nu+\frac{4-h}{2+h}\tau\pi+c.c.\bigr]+\frac{h}{4}\Bigl\{\frac{8+h}{2+h}(\tau\bar\tau+\pi\bar\pi)\Bigr\}
\end{multline}
\begin{multline}
\mathcal R_{3434}=\frac{1}{1+\omega}\Bigl[\nabla_{m\bar m}\omega-\frac{\nabla_l\omega\nabla_n\omega}{(1+\omega)(2+h)}-\frac{\nabla_m\omega\nabla_{\bar m}\omega}{1+\omega}+\frac{h}{2(2+h)}(\mu\nabla_l\omega-\rho\nabla_n\omega+c.c.)\Bigr]\\
+\frac{h}{2+h}\Bigl[\frac{(4+3h)}{4}(\mu-\bar\mu)(\rho-\bar\rho)+(\sigma\lambda-\mu\bar\rho+c.c.)\Bigr],
\end{multline}
\begin{multline}
\mathcal R_{1324}=-\frac{\nabla_{ln}\omega}{(1+\omega)(2+h)}+\frac{\nabla_{m\bar m}\omega}{2(1+\omega)}+\frac{\nabla_{m\bar m}h}{2(2+h)}+\frac{\nabla_l\omega\nabla_n\omega}{2(1+\omega)^2(2+h)}-\frac{\nabla_m\omega\nabla_{\bar m}\omega}{4(1+\omega)}\\
+\frac{\nabla_m\omega\nabla_{\bar m}h+c.c.}{4(1+\omega)(2+h)}-\frac{\nabla_mh\nabla_{\bar m}h}{4(2+h)^2}+\frac{h}{4(1+\omega)(2+h)}[\mu\nabla_l\omega-\rho\nabla_n\omega+(\pi-\bar\tau)\nabla_m\omega+c.c.]\\
+\frac{\bar\mu\nabla_lh-\bar\rho\nabla_nh}{2(2+h)}+\frac{1}{4(2+h)^2}\{[h\pi-(h+4)\bar\tau]\nabla_mh-[h\tau-(h+4)\bar\pi]\nabla_{\bar m}h\}\\
+\frac{h}{2(2+h)}\Bigl\{\nabla_{\bar m}\bar\pi-\nabla_n\bar\rho-\sigma\lambda-\kappa\nu+\bar\rho(\gamma+\bar\gamma)-\bar\pi(\alpha-\bar\beta)\\
-\frac{4+h}{4}(\rho-\bar\rho)(\mu-\bar\mu)-\mu\bar\rho-\frac{(4+h)(\tau\pi+\tau\bar\tau+\bar\tau\bar\pi)-h\pi\bar\pi}{2(2+h)}\Bigr\},
\end{multline}
\begin{multline}
\mathcal R_{1213}=-\frac{\nabla_{lm}\omega}{2(1+\omega)}-\frac{\nabla_{lm}h}{2(2+h)}+\frac{\nabla_lh\nabla_mh}{2(2+h)^2}+\frac{\nabla_l\omega}{4(1+\omega)}\Bigl[\frac{3\nabla_m\omega}{1+\omega}+\frac{\nabla_mh}{2+h}-\frac{h(\tau+\bar\pi)}{2+h}\Bigr]+\frac{h(\rho-\bar\rho)\nabla_m\omega}{8(1+\omega)}\\
-\frac{1}{2(2+h)}\Bigl\{\frac{(4+h)\bar\pi+2\tau}{2+h}\nabla_lh+\kappa\nabla_nh-\frac{h}{4}(\rho-\bar\rho)\nabla_mh\Bigr\}\\
-\frac{h}{2(2+h)}\Bigl\{\nabla_l(\tau+\bar\pi)-(\tau+\bar\pi)(\epsilon-\bar\epsilon)-\frac{2+h}{2}(\mu-\bar\mu)\kappa-2(\pi+\bar\tau)\sigma+\frac{h}{4}(\tau-\bar\pi)(\rho-\bar\rho)-(\rho+\bar\rho)\bar\pi-2\bar\rho\tau\Bigr\},
\end{multline}
\begin{multline}
\mathcal R_{1334}=-\frac{\nabla_{lm}\omega}{(1+\omega)(2+h)}+\frac{\nabla_l\omega}{2(1+\omega)(2+h)}\Bigl[\frac{3\nabla_m\omega}{1+\omega}+\frac{\nabla_mh}{2+h}-\frac{h(\tau+\bar\pi)}{2+h}\Bigr]+\frac{h(\rho-\bar\rho)\nabla_m\omega}{4(1+\omega)(2+h)}\\
+\frac{\sigma\nabla_{\bar m}h}{(2+h)^2}+\frac{\nabla_mh}{4(2+h)^2}\bigl[3h(\rho-\bar\rho)+4(\rho-2\bar\rho)\bigr]+\frac{h}{2(2+h)}\nabla_m(\rho-\bar\rho)\\
+\frac{h}{2(2+h)^2}\bigl[(\rho+\bar\rho)(\tau+\bar\pi)-2(\pi+\bar\tau)\sigma\bigr]-\frac{h}{4(2+h)}[(\rho-\bar\rho)(2\bar\alpha+2\beta+\tau-3\bar\pi)+4(\mu-\bar\mu)\kappa],
\end{multline}
\begin{multline}
\mathcal R_{1234}=-\frac{1}{2}\bigl\{(\bar\tau+\pi)\nabla_m\Bigl(\frac{h}{2+h}\Bigr)+\frac{1}{2+h}[(\mu\nabla_l-\rho\nabla_n)h+h(\beta-\bar\alpha+\nabla_m)(\bar\tau+\pi)+2h\sigma\lambda]-c.c.\bigr\},
\end{multline}
\begin{multline}
\mathcal R_{1313}=-h^2\sigma\nabla_l\Bigl(\frac{h}{2+h}\Bigr)-\frac{\kappa}{2+h}\nabla_mh\\
-\frac{h}{2(2+h)}\bigl[\nabla_l\sigma+\nabla_m\kappa+2(\rho-\bar\rho)\sigma+2(\tau+\bar\pi)\kappa-\sigma(\rho+\bar\rho+3\epsilon-\bar\epsilon)-\kappa(\tau-\bar\pi+\bar\alpha+3\beta)\bigr],
\end{multline}
\begin{multline}\label{R1314}
(2+h)\mathcal R_{1314}=-\frac{\nabla_{ll}\omega}{1+\omega}+\frac{3(\nabla_l\omega)^2}{2(1+\omega)^2}+\frac{\nabla_lh\nabla_l\omega}{(1+\omega)(2+h)}-\Bigl[\frac{\rho\nabla_lh}{2+h}+\frac{\bar\kappa\nabla_m[(1+\omega)h]}{2(1+\omega)}+c.c.\Bigr]\\
-\frac{h}{2}\Bigl[\nabla_l\bar\rho+\nabla_{\bar m}\kappa+\frac{4+h}{4}(\rho-\bar\rho)^2-\bar\rho^2-\sigma\bar\sigma-(\epsilon+\bar\epsilon)\bar\rho-\kappa(3\alpha+\bar\beta-2\pi-\bar\tau)+\bar\kappa\bar\pi\Bigr],
\end{multline}
\begin{multline}
\mathcal R_{1323}=\frac{1}{2(1+\omega)}\Bigl[\nabla_{mm}\omega-\frac{3(\nabla_m\omega)^2}{2(1+\omega)}\Bigr]+\frac{1}{2(2+h)}\Bigl[\nabla_{mm}h-\frac{(\nabla_mh)^2}{2(2+h)}-(\tau-\bar\pi)\nabla_mh\Bigr]\\
+\frac{\nabla_l[\bar\lambda(1+\omega)h]-\nabla_n[\sigma(1+\omega)h]}{2(1+\omega)(2+h)}-\frac{h}{2(2+h)}\Bigl[(2\mu+\bar\gamma-3\gamma)\sigma+(2\bar\rho+\epsilon-3\bar\epsilon)\bar\lambda+\frac{4+h}{2(2+h)}(\tau+\bar\pi)^2\Bigr].
\end{multline}
Here, `$c.c.$' means the complex conjugate of all proceeding terms in the same bracket. One is free to apply the Ricci identities corresponding to $R_{\mu\nu\alpha\beta}=0$ to the above results, e.g., the reality of $\mathcal R_{1314}$ can be shown using the Ricci identity involving $\nabla_l\rho-\nabla_{\bar m}\kappa$ and its complex conjugate.

\subsection{Vacuum Einstein's equations}


With $\mathcal R_{abcd}$ given, vacuum Einstein's equations $\tilde R_{ab}=0$ for the full metric are obtained using \eqref{Ricci_Riem}. In particular,
\begin{multline}
(1+\omega)(2+h)\tilde R_{12}=\\
\frac{2+h}{1+\omega}\nabla_{m\bar m}\omega+\nabla_{m\bar m}h-\frac{4\nabla_{ln}\omega}{1+\omega}-\frac{2\nabla_{ln}h}{2+h}+\frac{2\nabla_lh\nabla_nh}{(2+h)^2}+\frac{3\nabla_l\omega\nabla_n\omega}{(1+\omega)^2}+\frac{\nabla_m\omega\nabla_{\bar m}h+c.c.}{1+\omega}\\
+\frac1{2(1+\omega)}\Bigl\{(\mu\nabla_l-\rho\nabla_n)\bigl[(1+\omega)h\bigr]+\frac{\nabla_m\bigl[(1+\omega)^2(2+h)h(\pi-\bar\tau)\bigr]}{(1+\omega)(2+h)}+c.c.\Bigr\}\\
-\frac{(4+h)h}{4}\Bigl[(\rho-\bar\rho)(\mu-\bar\mu)-4\frac{\bar\tau\tau+\bar\pi\pi}{2+h}\Bigr]+\frac{h}{2}\Bigl[(\tau-\bar\pi)(\alpha-\bar\beta)-\frac{2h\tau\pi}{2+h}+c.c.\Bigr],
\end{multline}
\begin{multline}
\mathcal (1+\omega)\sqrt{2(2+h)}\tilde R_{13}=-\frac{1}{1+\omega}\Bigl[2\nabla_{lm}\omega-\frac{3\nabla_l\omega\nabla_m\omega}{1+\omega}\Bigr]-\frac{1}{2+h}\Bigl(\nabla_{lm}h-\frac{\nabla_lh\nabla_mh}{2+h}-\frac{\nabla_l\omega\nabla_mh}{1+\omega}\Bigr)\\
-h\frac{2(\tau+\bar\pi)\nabla_l\omega-(2+h)(\rho-\bar\rho)\nabla_m\omega}{2(1+\omega)(2+h)}-\frac{(4+h)\bar\pi+2\tau}{(2+h)^2}\nabla_lh-\frac{\kappa\nabla_nh}{2+h}+\Bigl[\frac{1+h}{2+h}\rho-\bar\rho\Bigr]\nabla_mh+\frac{\sigma\nabla_{\bar m}h}{2+h}\\
-\frac{h\nabla_l(\tau+\bar\pi)}{2+h}+\frac{h}{2}\nabla_m(\rho-\bar\rho)\\
+\frac{h}{2+h}\Bigl\{\frac{\rho-\bar\rho}{2}[(2\bar\pi-\tau)h-(\bar\alpha+\beta)(2+h)]+(\epsilon-\bar\epsilon)(\tau+\bar\pi)+3(\tau\bar\rho+\bar\pi\rho)+(\bar\tau+\pi)\sigma-\frac{1}{2}(2+h)(\mu-\bar\mu)\kappa\Bigr\},
\end{multline}
\begin{multline}
(1+\omega)\mathcal R_{34}=-\frac{2}{1+\omega}\Bigl(\nabla_{m\bar m}\omega-\frac{\nabla_{ln}\omega}{2+h}\Bigr)-\frac{\nabla_{m\bar m}h}{2+h}
+\frac{3\nabla_m\omega\nabla_{\bar m}\omega}{2(1+\omega)^2}+\frac{\nabla_mh\nabla_{\bar m}h}{2(2+h)^2}-\frac{\nabla_m\omega\nabla_{\bar m}h+c.c.}{2(1+\omega)(2+h)}\\
-\frac{\nabla_l\bigl[(\mu+\bar\mu)(1+\omega)^2h\bigr]-\nabla_n\bigl[(\rho+\bar\rho)(1+\omega)^2h\bigr]}{2(1+\omega)^2(2+h)}+\frac{(\bar\tau-\pi)\nabla_m\bigl[(1+\omega)h\bigr]+c.c.}{2(1+\omega)(2+h)}\\
+\frac{h}{2(2+h)}\Bigl\{\frac{4+h}{2+h}(\tau+\bar\pi)(\bar\tau+\pi)-(\mu-\bar\mu)(\rho-\bar\rho)h+4(\mu\bar\rho+\bar\mu\rho)-(\mu+\bar\mu)(\epsilon+\bar\epsilon)-(\rho+\bar\rho)(\gamma+\bar\gamma)\Bigr\}.
\end{multline}


\subsection{Petrov types}

The Petrov classification (cf.\ \cite{ChandrasekharBH,ExactSolns}) characterizes geometric properties of spacetimes according to the multiplicities of \emph{principal null directions} (PNDs). Under the tetrad formalism, $\Psi_0=0$ when $l$ is aligned with one of the PNDs, and the number of ways to (Lorentz-) transform an initially arbitrary tetrad to make $\Psi_0=0$ distinguishes different the Petrov types. More precisely, Lorentz transformations on tetrads (tetrad rotations) fall into three classes:
\begin{align}
\text{I: }&l\rightarrow l,\,n\rightarrow n+\bar am+a\bar m+a\bar al,\,m\rightarrow m+al\\
\text{II: }&n\rightarrow n,\,l\rightarrow l+\bar bm+b\bar m+b\bar bn,\,m\rightarrow m+bn\\
\text{III: }&l\rightarrow \lvert c\rvert l,\,n\rightarrow \lvert c\rvert^{-1}n,\,m\rightarrow c\lvert c\rvert^{-1}m,
\end{align}
for complex $a,b,c$. Weyl scalars transform as $\Psi_i=L_i^{\phantom{i}j}\Psi_j$, ($i,j=0\dotso4$) where
\begin{equation}
L_i^{\phantom{i}j}=
\begin{pmatrix}
1&&&&0\\\bar a&1&&\iddots&\\\bar a^2&2\bar a&1&&\\\bar a^3&3\bar a^2&3\bar a&1&\\\bar a^4&4\bar a^3&6\bar a^2&4\bar a&1
\end{pmatrix},
\begin{pmatrix}
1&4b&6b^2&4b^3&b^4\\&1&3b&3b^2&b^3\\&&1&2b&b^2\\&\iddots&&1&b\\0&&&&1
\end{pmatrix},
\mathrm{diag}[c^2,c,1,c^{-1},c^{-2}]
\end{equation}
for each class.
$\Psi_0$ can be made zero by a class II rotation:
\begin{equation}\label{Psi0=0}
\Psi_0\rightarrow\Psi_0+4b\Psi_1+6b^2\Psi_2^2+4b^3\Psi_3+b^4\Psi_4^4=0.
\end{equation}
Petrov types correspond to multiplicities of roots of \eqref{Psi0=0} for $b$: Type I (no multiple roots and PNDs), Type II (one double root and PND), Type D (two double roots and PNDs), Type III (one triple root and PND) and Type N (one quadruple root and PND). The specific criteria can be given \cite{ExactSolns} in terms of
\begin{align}
I&\equiv\Psi_0\Psi_4-4\Psi_1\Psi_3+3\Psi_2^2\\
J&\equiv2\Psi_1\Psi_2\Psi_3+\Psi_0\Psi_2\Psi_4-\Psi_2^3-\Psi_1^2\Psi_4-\Psi_0\Psi_3^2\\
K&\equiv\Psi_1\Psi_4^2-3\Psi_2\Psi_3\Psi_4+2\Psi_3^3\\
L&\equiv\Psi_2\Psi_4-\Psi_3^2,\quad N\equiv\Psi_4^2I-3L^2
\end{align}
as
\begin{align}
\text{Type I}&\quad\Leftrightarrow\quad I^3\neq27J^2\label{I}\\
\text{Type II}&\quad\Leftrightarrow\quad I^3=27J^2,(I,J,K,N\text{ not all vanishing})\label{II}\\
\text{Type D}&\quad\Leftrightarrow\quad I^3=27J^2,K=N=0,(I,J\text{ not both vanishing})\label{D}\\
\text{Type III}&\quad\Leftrightarrow\quad I^3=27J^2,I=J=0,(K,L\text{ not both vanishing})\label{III}\\
\text{Type N}&\quad\Leftrightarrow\quad I^3=27J^2,I=J=K=L=N=0\label{N}
\end{align}
(correcting a typo in the definition of $N$ in \cite{ExactSolns} according to e.g.\ \cite{Campanelli}). For the xKS form, one replaces $\Psi$'s with $\tilde\Psi$'s given by \eqref{Psi_R}:
\begin{multline}
4\sqrt2(1+\omega)(2+h)^{\frac{3}{2}}\tilde\Psi_1=\\
-\nabla_{lm}h+\frac{\nabla_lh\nabla_mh}{2+h}-\frac{[(4+h)\bar\pi+2\tau]\nabla_lh}{2+h}-\kappa\nabla_nh-\frac{(2+h)\rho-(4+h)\bar\rho}{2}\nabla_mh-\sigma\nabla_{\bar m}h\\
-h\nabla_l(\tau+\bar\pi)-\frac{h}{2}(2+h)\nabla_m(\rho-\bar\rho)\\
+\frac{h}{2}\bigl\{(2+h)[(\rho-\bar\rho)(\bar\alpha+\beta-\bar\pi)+3(\mu-\bar\mu)\kappa]+6(\pi+\bar\tau)\sigma+2(\tau+\bar\pi)(\bar\rho+\epsilon-\bar\epsilon)\bigr\}
\end{multline}
\begin{multline}
6(1+\omega)(2+h)\tilde\Psi_2=-\frac{2\nabla_{ln}h}{2+h}-\nabla_{m\bar m}h+\frac{2\nabla_lh\nabla_nh}{(2+h)^2}+\frac{\nabla_mh\nabla_{\bar m}h}{2+h}\\
+(\mu-2\bar\mu)\nabla_lh-(\rho-2\bar\rho)\nabla_nh+2\frac{\pi+2\bar\tau}{2+h}\nabla_mh-2\frac{\tau+2\bar\pi}{2+h}\nabla_{\bar m}h+[h\nabla_m(\pi+\bar\tau)-c.c.]\\
+h\bigl[(2+h)(\mu-\bar\mu)(\rho-\bar\rho)-\rho\bar\mu-(\gamma+\bar\gamma)\bar\rho+2\kappa\nu+\bar\kappa\bar\nu+4\lambda\sigma-\bar\lambda\bar\sigma\\
-(\pi+\bar\tau)(\bar\alpha-\beta)+(\tau+2\bar\pi)(\alpha-\bar\beta)+\tau\bar\tau+\frac{4(\tau+\bar\pi)(\bar\tau+\pi)}{(2+h)}\bigr].
\end{multline}

While the above criteria can be evaluated once an xKS solution is known, they are harder to impose when searching for new solutions, since they involve polynomials of $\tilde\Psi$'s. Simpler criteria can be given by noting that one can perform appropriate tetrad rotations so that \cite{ChandrasekharBH}
\begin{align}
&\tilde\Psi_0=\tilde\Psi_4=0\quad&&\text{for Type I}\label{specialPsi1I}\\
&\tilde\Psi_0=\tilde\Psi_4=\tilde\Psi_1=0\quad&&\text{for Type II}\\
&\tilde\Psi_0=\tilde\Psi_4=\tilde\Psi_1=\tilde\Psi_3=0\quad&&\text{for Type D}\\
&\tilde\Psi_0=\tilde\Psi_4=\tilde\Psi_1=\tilde\Psi_2=0\quad&&\text{for Type III}\\
&\tilde\Psi_0=\tilde\Psi_1=\tilde\Psi_2=\tilde\Psi_3=0\quad&&\text{for Type N}.\label{specialPsi1N}
\end{align}
The simultaneous vanishing of $\tilde\Psi_0,\tilde\Psi_4$ in the first four types means that $\tilde l,\tilde n$ are each aligned with a PND. Note also that we have the symmetry under the interchanges $\tilde l\leftrightarrow\tilde n,\tilde m\leftrightarrow\bar{\tilde m},\tilde\Psi_0\leftrightarrow\tilde\Psi_4,\tilde\Psi_1\leftrightarrow\tilde\Psi_3$. In the equivalent treatment of Petrov classification as an eigenvalue problem on bivectors, the Weyl principal tetrad is introduced for which $\tilde\Psi$'s take another set of special values \cite{ExactSolns}:
\begin{align}
&\tilde\Psi_0=\tilde\Psi_4,\tilde\Psi_1=\tilde\Psi_3=0\quad&&\text{for Type I}\label{specialPsi21I}\\
&\tilde\Psi_0=\tilde\Psi_1=\tilde\Psi_3=0,\tilde\Psi_4=-2\quad&&\text{for Type II}\\
&\tilde\Psi_0=\tilde\Psi_4=\tilde\Psi_1=\tilde\Psi_3=0\quad&&\text{for Type D}\\
&\tilde\Psi_0=\tilde\Psi_4=\tilde\Psi_1=\tilde\Psi_2=0,\tilde\Psi_3=-i\quad&&\text{for Type III}\\
&\tilde\Psi_0=\tilde\Psi_1=\tilde\Psi_2=\tilde\Psi_3=0\quad&&\text{for Type N}.\label{specialPsi2N}
\end{align}
Of course the special tetrad choices in \eqref{specialPsi1I}--\eqref{specialPsi2N} do not necessarily coincide with the ones in which the metric admits xKS form. We proceed by studying some special cases for the background tetrad and spin coefficients.

\subsection{Special choices for flat background tetrad and spin coefficients}

For the flat metric $ds^2=-(\omega^0_\mu dx^\mu)^2+(\omega^1_\mu dx^\mu)^2+(\omega^2_\mu dx^\mu)^2+(\omega^3_\mu dx^\mu)^2$ written using some orthonormal basis $\omega^a_\mu$, the null tetrad can be constructed as $l,n=\frac{\omega^0\pm\omega^1}{\sqrt2}$, $m,\bar m=\frac{\omega^2\pm i\omega^3}{\sqrt2}$. Consider e.g.\ the metric in Cartesian coordinates with $l,n=\frac{-dt\pm dx}{\sqrt2}$, $m,\bar m=\frac{dy\pm idz}{\sqrt2}$, for which all the spin coefficients, and consequently $\tilde\Psi_0$ \& $\tilde\Psi_4$, vanish. Then by \eqref{I}--\eqref{N} the following subcases for the Petrov types are possible:
\begin{align}
\tilde\Psi_1=0,\tilde\Psi_2=0,\tilde\Psi_3=0:\quad&\text{N}&\tilde\Psi_1\ne0,\tilde\Psi_2=0,\tilde\Psi_3=0:\quad&\text{N}\label{Psi123_1}\\
\tilde\Psi_1=0,\tilde\Psi_2=0,\tilde\Psi_3\ne0:\quad&\text{III}&\tilde\Psi_1\ne0,\tilde\Psi_2=0,\tilde\Psi_3\ne0:\quad&\text{I}\\
\tilde\Psi_1=0,\tilde\Psi_2\ne0,\tilde\Psi_3=0:\quad&\text{D}&\tilde\Psi_1\ne0,\tilde\Psi_2\ne0,\tilde\Psi_3=0:\quad&\text{D}\\
\tilde\Psi_1=0,\tilde\Psi_2\ne0,\tilde\Psi_3\ne0:\quad&\text{II}&\tilde\Psi_1\ne0,\tilde\Psi_2\ne0,\tilde\Psi_3\ne0:\quad&\text{I or II}.\label{Psi123_8}
\end{align}
Now $\tilde\Psi_{1,2,3}$ depend only on $\omega,h$ which are solutions to $\tilde R_{ab}=0$. For the current tetrad choice, one simply evaluate the directional derivatives in $\tilde R_{ab}$ using partial derivatives, e.g.\ $\nabla_{ln}h=l^\mu n^\nu\partial_\mu\partial_\nu h$, etc.. We find the following solutions for $\omega=\omega(t,x),h=h(t,x)$, defining new variables $u\equiv t-x,v\equiv t+x$,
\begin{align*}
\omega(u,v)&=\omega(u),\;\chi^2(u,v)=\frac{\omega'(u)V(v)}{\sqrt{1+\omega(u)}}\\
\omega(u,v)&=\omega(v),\;\chi^2(u,v)=\frac{\omega'(v)U(u)}{\sqrt{1+\omega(v)}}\\
\omega(u,v)&=U(u)+V(v),\;\chi^2(u,v)=\pm\frac{U'(u)V'(v)}{\sqrt{1+U(u)+V(v)}}
\end{align*}
where $U,V$ are arbitrary functions and $\chi^2=(1+\omega)(1+h/2)$. The first two solutions make all $\tilde\Psi$'s vanish and thus are flat spacetimes. The third solution yields a non-vanishing $\tilde\Psi_2=-\frac{1}{2}(1+\omega)^{-\frac{3}{2}}$ (type D) and the metric becomes
\begin{equation}
ds^2=\mp\frac{dUdV}{\sqrt{1+U+V}}+(1+U+V)(dy^2+dz^2)
\end{equation}
treating $U,V$ as new coordinates. It can be transformed into the `AIII metric' introduced by Ehlers and Kundt (see \S9.1.2 of \cite{GriffithsPodolsky} for details):
\begin{equation}\label{AIII}
ds^2=\pm(r^{-1}d\tau^2-rdr^2)+r^2(dy^2+dz^2),
\end{equation}
using $U=r^2/2-\tau-1$, $V=r^2/2+\tau$.

As a second example, consider the flat metric in oblate spheroidal coordinates \cite{Teukolsky:Kerr}:
\begin{align}
ds^2&=-dt^2+\frac{\Sigma}{r^2+a^2}dr^2+\Sigma d\theta^2+(r^2+a^2)\sin^2\theta d\phi^2,\quad(\Sigma\equiv r^2+a^2\cos^2\theta)\label{flat_Kerr0}\\
&=-\frac{r^2+a^2}{\Sigma}(dt-a\sin^2\theta d\phi)^2+\frac{\Sigma}{r^2+a^2}dr^2+\Sigma d\theta^2+\frac{\sin^2\theta}{\Sigma}[(r^2+a^2)d\phi-adt]^2\label{flat_Kerr}\\
&\overset{a=0}{=\joinrel=}-dt^2+dr^2+r^2d\theta^2+r^2\sin^2\theta d\phi^2.\label{flat}
\end{align}
The second line \eqref{flat_Kerr} generates the Kerr metric after multiplying the first term by $1-2mr/(r^2+a^2)$ and dividing the second term by the same factor; similarly, the third line \eqref{flat} generates the Schwarzschild metric after multiplying the first term by $1-2m/r$ and dividing the second term by the same factor. They both yield $\tilde\Psi_0=\tilde\Psi_4=0$, so the same possibilities \eqref{Psi123_1}--\eqref{Psi123_8} arise. The first line \eqref{flat_Kerr0}, on the other hand, correspond to general, all non-vanishing $\Psi$'s. Lastly, consider the flat metric in bispherical coordinates used to construct wormhole initial conditions \cite{Misner}
\begin{equation}\label{key}
ds^2=-dt^2+(\cosh\mu-\cos\theta)^{-2}[d\mu^2+(d\theta^2+\sin^2\theta d\phi^2)],\quad\pi<\mu\le\pi.
\end{equation}
Again, all $\Psi$'s are generally non-vanishing. Solving Einstein's equations for these cases is left for future work.

\section{Comparison with double-null foliation}

We draw connections between the xKS form and another `(2+2)-split' ansatz, namely the double-null foliation \cite{Hayward,Werner,Vickers:doublenull}, and show that one can identify them in special gauges. We temporarily drop tilde for the double-null foliation formulae below. Following derivations in \cite{Werner}, one foliate the spacetime by introducing two null hypersurfaces $u^A(x^\alpha)=\text{const.},(A=0,1)$ and which intersect in spacelike 2-surfaces with intrinsic coordinates $\theta^a(x^\alpha),(a=2,3)$. Define normals to the null hypersurfaces and tangent vectors to the spacelike 2-surfaces respectively as
\begin{equation}
k^{(A)}_\alpha\equiv e^\lambda\partial_\alpha u^A,\quad e^\alpha_{(a)}\equiv\frac{\partial x^\alpha}{\partial\theta^a}
\end{equation}
for some function $\lambda$.
One has the conditions
\begin{equation}\label{norm_l_e}
g^{\alpha\beta}k^{(A)}_\alpha k^{(B)}_\beta=e^\lambda\eta^{AB},\quad g_{\alpha\beta}e^\alpha_{(a)}e^\beta_{(b)}=h_{ab},\quad k^{(A)}_\alpha e_{(a)}^\alpha=0
\end{equation}
and the completeness relations
\begin{align}
g_{\alpha\beta}&=e^{-\lambda}\eta_{AB}k^{(A)}_\alpha k^{(B)}_\beta+h_{ab}e^{(a)}_\alpha e^{(b)}_\beta\\
g^{\alpha\beta}&=e^{-\lambda}\eta^{AB}k_{(A)}^\alpha k_{(B)}^\beta+h^{ab}e_{(a)}^\alpha e_{(b)}^\beta,
\end{align}
where $\eta_{AB}=\eta^{AB}=\text{anti-diag}(-1,-1)$ and $h_{ab}$ (with inverse $h^{ab}$) are intrinsic metrics on subspaces $(u^0,u^1)$ and $(\theta^2,\theta^3)$ respectively. The coordinate basis one-forms and vectors are expressed in terms of those of intrinsics coordinates as
\begin{align}
dx^\alpha&=k^\alpha_{(A)}du^A+e^\alpha_{(a)}(d\theta^a+s^a_Adu^A)\label{dx}\\
\partial_\alpha&=e^{(a)}_\alpha\partial_a+e^{-\lambda}k^{(A)}_\alpha(\partial_A-s^a_A\partial_a)\label{px}
\end{align}
where $s_A^a=e^{(a)}_\alpha\frac{\partial x^\alpha}{\partial u^A}$ are the `shifts' of tangent vectors to $u^A$ along $\theta^a$ directions. We then arrive at the (2+2)-split of the metric
\begin{align}
g_{\alpha\beta}dx^\alpha dx^\beta&=e^\lambda\eta_{AB}du^Adu^B+h_{ab}(d\theta^a+s^a_Adu^A)(d\theta^b+s^b_Bdu^B)\label{gdd}\\
g^{\alpha\beta}\partial_\alpha\partial_\beta&=h^{ab}\partial_a\partial_b+e^{-\lambda}\eta^{AB}(\partial_A-s^a_A\partial_a)(\partial_B-s^b_B\partial_b).
\end{align}

It was shown \cite{Werner,Vickers:doublenull} that gauge choices can be made so that $s_A^a$ vanish, in which case it is easier to bring the 2-d metric $h_{ab}d\theta^ad\theta^b$ to a conformally flat form $W(x,y)^2(dx^2+dy^2)$ in the so-called isothermal coordinates \cite{Chern}. Then the metric \eqref{gdd} is in exactly the xKS form:
\begin{equation}\label{doublenull_xKS}
ds^2=-2e^\lambda du^0du^1+W(x,y)^2(dx^2+dy^2)
\end{equation}
upon identifying $e^\lambda=\chi^2$, where $(u^0,u^1,x,y)$ serve as coordinates for the flat background.

A useful quantity defined in the double-null foliation is the expansion \cite{Hayward}
\begin{equation}\label{expan}
\theta=\frac{1}{2}h^{\alpha\beta}\mathcal L_Nh_{\alpha\beta}=h^{\alpha\beta}N_{\alpha;\beta}
\end{equation}
where $\mathcal L_N$ is the Lie derivative along $N^\alpha\equiv g^{\alpha\beta}k^{(B)}_\beta\eta_{AB}$ assumed to be future outgoing \footnote{Note that $e^{\lambda}$ here equals $e^{-f}$ in \cite{Hayward}, and $(n_\pm)_\alpha$ there should be identified with $e^{-\lambda}k^{(B)}_\alpha\eta_{AB}$ here due to the sign conventions.}. The \emph{trapping horizon} lies at $\theta=0$. We translate this condition to the xKS form. According to \eqref{doublenull_xKS}, we choose the future outgoing background tetrad vector to be, restoring our convention about tilde,
\begin{equation}
l^\alpha=\partial x^\alpha/\partial u^0=\tilde k^\alpha_{(0)},\quad l_\alpha=-\partial_\alpha u^1=-e^{-\lambda}\tilde k^{(1)}_\alpha.
\end{equation}
Then $\tilde N^\alpha=l^\alpha,\tilde N_\alpha=e^\lambda l_\alpha$, and the two expressions for $\theta$ in \eqref{expan} give identical results:
\begin{equation}\begin{aligned}
\theta&=\frac{1}{2}(1+\omega)^{-1}(m^\alpha\bar m^\beta+m^\beta\bar m^\alpha)\tilde{\mathcal L}_{l}[(1+\omega)(m_\alpha\bar m_\beta+m_\beta\bar m_\alpha)]\\
&=(1+\omega)^{-1}(m^\alpha\bar m^\beta+m^\beta\bar m^\alpha)\tilde\nabla_\alpha(e^\lambda l_\alpha)\\
&=\frac{\nabla_l\omega}{1+\omega}-\rho-\bar\rho.
\end{aligned}\end{equation}

\section{Conclusion}

We have rewritten basic contents of general relativity in the xKS form, including Riemann tensor, vacuum Einstein's equations and criteria for algebraic types of the spacetime. These involve up to second order derivatives of only two free functions $W,h$, with all remaining pieces determined by the choice of a background tetrad. Hopefully this could provide new angles to address the non-linearity of Einstein's equations and to find new solutions. The fact that the background tetrad can be transformed into the full metric tetrad simply by scalings makes the xKS form suitable to discuss algebraic types of the full spacetime. It turns out the such `xKS transformations' (between two sets of tetrads) does not commute with the Lorentz rotations on either tetrad, which means that the full metric is really in a preferred gauge to assume the xKS form, and also that by applying a Lorentz rotation first to the background tetrad, the xKS transformation would lead to a different full metric. One may also generalize the xKS form to include the energy-momentum tensor, e.g., that of an electromagnetic field, with the vector potential given, say, by the linear combinations of the tetrad. We leave this for future work.



\bibliography{xKS-bib}

\end{document}